\documentclass{ws-rv9x6}
\usepackage{ws-rv-thm}
\usepackage{ws-rv-van}
\usepackage{hyperref}

\usepackage{subfigure}
\usepackage{caption}
\usepackage{float}
\restylefloat{table}
\restylefloat{figure}

\def\be{\begin{equation}}
\def\ee{\end{equation}}
\def\bea{\begin{eqnarray}}
\def\eea{\end{eqnarray}}
\def\ov{\overline}

\allowdisplaybreaks[2]
\numberwithin{equation}{section}

\begin{document}


\chapter[Towards Systematics of Calabi-Yau Landscape for String Cosmology]{Towards Systematics of Calabi-Yau Landscape \\for String Cosmology
\label{ra_ch1}}

\author[G.~K.~Leontaris and P.~Shukla]{George K. Leontaris$^\dagger$ and Pramod Shukla$^\ddagger$\footnote{Speaker at BIRS-CMI workshop ``Recent Progress in Computational String Geometry" held during January 26-31, 2026 at Chennai Mathematical Institute (CMI), Chennai.}}

\address{$^\dagger$Physics Department, University of Ioannina\\
	University Campus, Ioannina 45110, Greece \\
Email: leonta@uoi.gr\\
\vskip0.2cm 
$^\ddagger$Department of Physical Sciences, Bose Institute,\\
	Unified Academic Campus, EN 80, Sector V, \\ Bidhannagar, Kolkata 700091, India\\
Email: shukla@jcbose.ac.in
}

\begin{abstract}
In this review, we discuss the relevance and impact of studying Calabi-Yau threefolds in the context of global model building in string phenomenology. First, taking a phenomenologist-friendly approach, we review how the topologies of the various divisors and curves of the compactifying CY threefolds play a crucial role for generating the various ``suitable" classes of effective scalar potentials, within the framework of the popular moduli stabilization schemes such as KKLT and LVS. Subsequently, we will discuss the impact of the specifics of the CY threefold geometries in the minimal LVS inflationary models such as fibre inflation, in particular, along the challenges such as the inflaton field-range bound. In this regard, we discuss a multi-field approach in which several fibre moduli assist to drive successful inflation having a sufficient number of efolds, without getting close to their individual K\"ahler cone boundaries.
\end{abstract}

\section{Introduction}
\label{sec:intro}

Finding the ``right" Calabi Yau (CY) threefold has been among the central tasks for constructing four-dimensional (semi)realistic models using superstring compactifications. Being at the core of model building, this task has resulted in an enormous amount of efforts aiming at constructing and classifying CY threefolds since more than three decades \cite{Green:1986ck, Candelas:1987kf, Green:1987cr, Candelas:1993dm, Batyrev:1993oya,Candelas:1994hw,Hosono:1994ax,Kreuzer:2000xy,Gray:2013mja}. At present we have two main CY threefold datasets, namely the Complete Intersection CY threefolds (CICYs) and Toric Hypersurface CY threefolds (THCYs). The projective CICY database of ~\cite{Candelas:1987kf,Anderson:2017aux} has been also refereed as ``pCICY" database as there are additional so-called ``generalized" CICYs referred as ``gCICY" \cite{Anderson:2015iia} along with another possible dataset of the Toric CICY refereed as ``tCICY"; and a couple of examples of tCICYs have been presented in \cite{Cicoli:2021dhg}.

The pCICYs are realized as multi-hypersurfaces in the product of projective spaces \cite{Green:1986ck}, and a complete list of 7890 such CICYs along with their Hodge numbers $(h^{1,1}, h^{2,1})$ have been presented in  \cite{Candelas:1987kf,Green:1987cr}. Moreover, the fibration structures of CICYs have been studied recently in \cite{Gray:2014fla,Anderson:2017aux} with some initiatives taken for constructing local MSMS-like models \cite{Anderson:2011ns, Anderson:2012yf, Anderson:2013xka},  and moduli stabilization in Heterotic/type-IIB based setups \cite{Anderson:2010mh,Anderson:2011cza, Bobkov:2010rf,Carta:2021sms,Carta:2021uwv, Carta:2022web,Carta:2022oex}. Further, a complete list of explicit classification of CICY orientifolds with non-trivial $(1,1)$-cohomology has been also presented in \cite{Carta:2020ohw}, while a thorough analysis and classification of the (toric) divisor topologies have been presented in \cite{Carta:2022web}. In the meantime, a complete list of CICY fourfolds has been also developed \cite{Gray:2013mja}. It is worth mentioning that for pCICYs, most of the topological information, such as triple intersection numbers, second Chern class, Euler characteristics, Mori cone, K\"ahler cone and even the Gopakumar-Vafa invariants \cite{Gopakumar:1998ii,Gopakumar:1998jq}, needed for model building have been computed in a series of work \cite{Anderson:2017aux,Hosono:1994ax,Carta:2021sms}.

The THCYs are realized as hypersurfaces in toric varieties \cite{Batyrev:1993oya} arising from appropriate triangulations of the reflexive polytopes classified in the Kreuzer-Skarke (KS) database \cite{Kreuzer:2000xy}. Computing the topological properties of some of the THCYs have been initiated in early nineties \cite{Candelas:1993dm, Candelas:1994hw,Hosono:1994ax}, however a huge surge in systematically exploring the KS database CYs have been witnessed in recent years due to some efficient tools/packages being developed. These include, for example, the ``Package for analyzing lattice polytopes" (PALP) \cite{Kreuzer:2002uu}, its updated new offspring version \cite{Braun:2011ik} with \texttt{mori.x} module, SAGE \cite{sagemath}, \texttt{cohomCalg} package  \cite{Blumenhagen:2010pv,Blumenhagen:2011xn}, and \href{https://cytools.liammcallistergroup.com/about/}{\texttt{CYTools}}. 

In  the meantime, a sort of {\it phenomenologist-friendly} dataset for THCYs has been developed in \cite{Altman:2014bfa} which is referred as Altman-Gray-He-Jejjala-Nelson (AGHJN) dataset. Developed for CYs with $1\leq h^{1,1}(CY) \leq 6$, this is an amazing data collection equipped with mostly all the necessary model building first-hand information, e.g. GLSM data, SR ideal, Fundamental group for defining CY threefold, Hodge numbers of the CY threefold, Chern classes, Triple intersection numbers, Mori/K\"ahler cone etc. This database has been subsequently used for constructing models \cite{Altman:2017vzk, Cicoli:2018tcq}. Moreover, this AGHJN-dataset has been updated with odd-orientifold constructions for CY threefolds in \cite{Altman:2021pyc} as an extension of the previous work in \cite{Gao:2013pra}. In addition, further extension of CY orientifold dataset of \cite{Altman:2014bfa, Altman:2021pyc} with the inclusion of THCYs with $h^{1,1}(CY) \geq 7$ has been initiated in \cite{Gao:2021xbs, Crino:2022zjk}. 

Which CY threefold is better or  more suitable for realistic string model building has been one of the prime questions still seeking an answer up to a satisfactory extent. However one can always make a classification for the known lists of CY threefolds to suggest that ``some" class of CY threefold can be more useful for certain purposes as compared to the other ones. In this regard, knowing divisor topologies of the compactifying CY threefolds in a systematic way can be a useful step towards equipping one with ingredients needed for performing a classification for global model building. In recent years, there have been huge amount of efforts in this field of computational CY geometries, especially with the advent of the \href{https://cytools.liammcallistergroup.com/about/}{\texttt{CYTools}} which has been proven to be very efficient in triangulating the polytopes, including those corresponding to larger $h^{1,1}$. Moreover, it has been also demonstrated to perform divisor topology computations, and its subsequent phenomenological implications have been initiated in \cite{Braun:2017nhi,Demirtas:2018akl,Demirtas:2020dbm}.

The structure of compactifying CY has a huge impact on the model building as the internal geometries such as divisor and curve topologies, involutions, brane-setting, tadpole cancelation conditions, K\"ahler cone conditions, etc. play a crucial role in determining the effective scalar potential of the four-dimensional supergravity theory. {\it In this review, our central focus will be to briefly highlight the impact of computing/classifying the divisor topologies of the CY threefolds on the broader goal of constructing global realistic models in CY superstring compactification frameworks.}


\section{Need For Global Model Building}
\label{sec:intro}


While building a model, say a model of inflationary cosmology within string theoretic framework, one considers several assumptions (e.g. in the form of Ansatz for the K\"ahler potential ${\cal K}$ and and superpotential $W$) and it is necessary to look for explicit constructions in which those assumptions are truly fulfilled. Finding the ``right" compactifying CY threefold which can result in `appropriate' and `just-enough' corrections is the core of looking for a global inflationary models. Both these conditions are necessary; e.g. the holomorphic involution and brane-setting should not only be rich enough to induce appropriate corrections but should also avoid having the unwanted or too many contributions to the effective scalar potential, in order to maintain the flatness of the inflationary potential. 

\subsection{Moduli Stabilization Schemes}
The F-term scalar potential governing the low energy dynamics of the four-dimensional ${\cal N} = 1$ effective supergravity arising from the type IIB superstring compactifications is encoded in 
\bea
V = e^{{\cal K}}({\cal K}^{{\cal A} \ov {\cal B}} (D_{\cal A} W) (D_{\ov {\cal B}} \ov{W}) -3 |W|^2),
\eea
where the K\"ahler potential (${\cal K}$) and the holomorphic superpotential ($W$)  depend on the various moduli, e.g. the complex-structure (CS) moduli $U^i$,  the axio-dilaton $S$ and the K\"ahler moduli $T_\alpha$. For a decoupled structure of ${\cal K}$ such that ${\cal K} \equiv {\cal K}_{\rm cs} (U^i)+ {\cal K}(S, T_\alpha)$, the inverse  metric is block-diagonal and $V \equiv V_{\rm cs} + V_{\rm k}$ such that $V_{\rm cs} =  e^{{\cal K}} {\cal K}_{\rm cs}^{i \ov {j}} (D_i W) (D_{\ov {j}} \ov{W})$ and $V_{\rm k} = e^{{\cal K}}({\cal K}^{{A} \ov {B}}(D_{A} W)(D_{\ov {B}} \ov{W}) -3 |W|^2)$ where $A, B \in \{S, T_\alpha\}$. Moduli stabilisation follows a two-step methodology. First, one fixes the CS moduli $U^i$ and the axio-dilaton $S$ by the leading order flux superpotential $W_{\rm flux}$ induced by the 3-form fluxes $(F_3, H_3)$ \cite{Gukov:1999ya} via the  supersymmetric flatness conditions $D_{U^i} W_{\rm flux} = 0 = D_{S} W_{\rm flux}$.

After achieving the supersymmetric stabilisation of the axio-dilaton and the CS moduli, one has $\langle W_{\rm flux} \rangle = W_0$. However, the no-scale structure protects the K\"ahler moduli $T_\alpha$ which remain flat, and as a second step, they can be stabilised via including other sub-leading contributions, e.g. from the various possible (non-)perturbative corrections arising from the whole series of $\alpha^\prime$ and string-loop ($g_s$) corrections. At two-derivative order, these effects can be captured through corrections in ${\cal K}$ and $W$,
\bea
\label{eq:KandW}
& & {\cal K} = {\cal K}_{\rm cs} -\ln\left[-\,i\,(S-\bar{S})\right]-2\ln{\cal Y}, \\
& & W = W_0 + W_{\rm np}(S, T_\alpha), \nonumber
\eea
where ${\cal K}_{\rm cs}$ depends on CS moduli, while ${\cal Y}$ generically encodes the volume/dilaton pieces such that ${\cal Y} = {\cal V}$ at tree-level. Here ${\cal V} = \frac{1}{3!} k_{\alpha\beta\gamma} t^\alpha t^\beta t^\gamma$ where $k_{\alpha\beta\gamma}$ is the triple intersection number of the CY$_3$ and the 2-cycle volumes $t^\alpha$ is related to the divisor volume $\tau_\alpha$ via $\tau_\alpha = \partial_{t^\alpha} {\cal V}$.

\vskip0.2cm
\noindent
{\bf Standard LVS:} The LVS scheme of moduli stabilisation considers a combination of perturbative $(\alpha^\prime)^3$ corrections to ${\cal K}$ \cite{Becker:2002nn} and a non-perturbative contribution to $W$ \cite{Witten:1996bn}. The minimal LVS construction includes two K\"ahler moduli with a so-called Swiss-cheese like volume form: ${\cal V} = \lambda_b \tau_{b}^{3/2} - \lambda_s \tau_{s}^{3/2}$ where $\lambda_{b} = {\sqrt2}/({3\sqrt{k_{bbb}}})$ and $\lambda_s ={\sqrt2}/({3\sqrt{k_{sss}}})$. The K\"ahler potential ${\cal K}$ is given by Eq.~(\ref{eq:KandW}) with ${\cal Y} = {\cal V}+{\xi}/({2 g_s^{3/2}})$ such that $\xi=-{\chi({\rm CY}_3)\,\zeta(3)}/({2\,(2\pi)^3})$, and $\chi({\rm CY}_3)$ being the CY Euler characteristic and $\zeta(3)\simeq 1.202$. It is worth recalling that moduli stabilisation in LVS requires $\xi>0$ or equivalently $\chi<0$, i.e. a constraint $h^{1,1}< h^{2,1}$ on the Hodge numbers counting the K\"ahler- and CS-moduli.

Furthermore, non-perturbative superpotential contributions \cite{Witten:1996bn} can be generated by using rigid divisors, such as shrinkable del-Pezzo (dP) 4-cycles, or by rigidifying non-rigid divisors using magnetic fluxes \cite{Bianchi:2011qh, Bianchi:2012pn, Louis:2012nb}. In fact, the presence of a `diagonal' del-Pezzo (ddP) divisor, the so-called `small' $4$-cycle of the CY$_3$ gives: $W= W_0 + A_s\, e^{- i\, a_s\, T_s}$, where $T_s = c_s - i \tau_s$ for $c_s$ is the $C_4$ axion, and the holomorphic pre-factor $A_s$ is considered as a parameter after fixing CS-moduli. Subsequently, after minimising the axion $c_s$, the leading order pieces in the large volume expansion are collected as \cite{Balasubramanian:2005zx}:
\bea
\label{eq:VlvsSimpl}
& & \hskip-0.5cm V_{\rm LVS} \simeq \frac{\alpha_1}{{\cal V}^3} - \,\frac{\alpha_2\tau_s}{{\cal V}^2}\, e^{- a_s \tau_s} +\frac{ \alpha_3\,\sqrt{\tau_s}}{{\cal V}}\, e^{-2 a_s \tau_s},
\eea
where $\alpha_i$'s are positive constants defined as $\alpha_1 = \frac{3}{4} \kappa\hat\xi |W_0|^2$, $\alpha_2 = 4 \kappa a_s |W_0| |A_s|$, $\alpha_3 = 4 \kappa a_s^2 |A_s|^2 \sqrt{2 k_{sss}}$, while $\kappa = g_s e^{K_{\rm cs}}/(8 \pi)$. This potential (\ref{eq:VlvsSimpl}) results in exponentially large $\langle{\cal V}\rangle$ determined by
\bea
\label{eq:}
& & \hskip-0.75cm \boxed{\langle {\cal V} \rangle \simeq \frac{\alpha_2 \sqrt{\langle \tau_s \rangle}}{2\,\alpha_3}\, e^{a_s \langle \tau_s \rangle}, \qquad \langle \tau_s \rangle \simeq \hat\xi^{2/3} \, \left(\frac{9\,k_{sss}}{8}\right)^{1/3}.}
\eea
Here $k_{sss} = 9 - n$ is the degree of the dP$_n$ divisor such that dP$_0 = {\mathbb P}^2$. Thus the LVS models based on Swiss-cheese CY with $h^{1,1}({\rm CY}) \geq 3$ have leading order flat directions which can act as promising inflaton candidate with a sub-leading potential. However, the K\"ahler cone constraints are crucially important, e.g.~in large field models like Fibre inflation.

\vskip0.2cm 
\noindent
{\bf Perturbative LVS:} With the inclusion of the so-called {\it log-loop} effects \cite{Antoniadis:2018hqy,Antoniadis:2018ngr,Antoniadis:2019doc,Antoniadis:2019rkh,Antoniadis:2020ryh,Antoniadis:2020stf} along with the BBHL corrections \cite{Becker:2002nn}, the K\"ahler potential ${\cal K}$ is given by Eq.~(\ref{eq:KandW}) with ${\cal Y} ={\cal V} +  \frac{\xi}{2} \, e^{-\frac{3}{2} \phi} + e^{\frac{1}{2} \phi}\, \left(\sigma + \eta \, \ln{\cal V}\right)$, where $\xi = - \frac{\chi({\rm CY})\, \zeta[3]}{2(2\pi)^3}, \,\sigma  = - \frac{\chi({\rm CY})\, \zeta[2]}{2(2\pi)^3} \sigma_0, \,\eta =  \frac{\chi({\rm CY})\, \zeta[2]}{2(2\pi)^3} \eta_0, \,\frac{\xi}{\eta} = -\frac{\zeta[3]}{\zeta[2]} $. We note that, the parameters $\xi$, $\sigma$ and $\eta$ do not depend on the string coupling $g_s$ following from the $SL(2, {\mathbb Z})$ arguments, and should be fixed after first step of moduli stabilization \cite{Leontaris:2022rzj}. However, $\sigma$ and $\eta$ parameters generically may have a dependence on the complex structure moduli, and for that purpose we have introduced two new parameters, namely $\sigma_0$ and $\eta_0$, while still keeping the $SL(2,\mathbb Z)$ motivated factors appearing with the Riemann $\zeta$-functions and the Euler characteristic of the CY threefold.

Now, considering the tree-level superpotential $W_0$ for CS+dilaton moduli stabilisation, one gets the following scalar potential \cite{Leontaris:2022rzj},
\bea
\label{eq:pheno-potV2}
& & \hskip-0.5cm V_{\rm pLVS} =  \frac{3\kappa |W_0|^2}{4{\mathcal V}^3}\left(\hat\xi + 2\,\hat\eta \, \ln{\cal V} - 8\,\hat\eta + 2\,\hat\sigma \right).
\eea
This subsequently results in an exponentially large VEV for the overall volume modulus determined as,
\bea
\label{eq:pert-LVS}
& & \hskip-0.5cm \langle {\cal V} \rangle \simeq e^{\frac{13}{3}-\frac{\hat\xi}{2\, \hat\eta} -\frac{\hat\sigma}{ \hat\eta}} = e^{a/g_s^2 + b}, \quad  a = \frac{\zeta[3]}{2 \zeta[2]\eta_0} \simeq \frac{0.365381}{\eta_0}, \quad b = \frac{13}{3}+\frac{\sigma_0}{\eta_0}~\cdot \nonumber
\eea
For natural values $\sigma_0 = -2$ and $\eta_0 = 1$, the numerical estimate for $g_s = 0.2$ gives $\langle {\cal V} \rangle = 95594.5$ while $g_s = 0.1$ leads to $\langle {\cal V} \rangle = 7.615 \cdot 10^{16}$. Further, similar to the standard Swiss-Cheese LVS case, this corresponds to a non-supersymmetric AdS minimum. Given that an exponentially large VEV of the overall volume $\cal V$ is obtained by using only the perturbative effects, this scheme is referred as ``perturbative LVS".

\subsection{On Generating `Suitable' Effective Scalar Potential}
As we motivated earlier, knowing divisor topologies of the compactifying CY threefolds in a systematic way can be a useful step towards equipping one with ingredients needed for performing a classification for global model building. In fact, exploring suitable geometries can help in knowing the structure of induced scalar potential, and hence can facilitate/dictate moduli stabilization and any subsequent phenomenological applications such as inflation or post-inflationary issues. Let us elaborate more on these points:

\vskip0.1cm
\noindent
{\bf (i)~Non-perturbative superpotential:} In order to generate non-perturbative superpotential contributions which are central ingredients for K\"ahler moduli stabilization schemes, e.g. \cite{Kachru:2003aw,Balasubramanian:2005zx}, one needs suitable four-cycles with unit Arithmetic genus \cite{Witten:1996bn}, (see \cite{Blumenhagen:2010ja} also for zero-mode analysis). In this regard, rigid divisors (in particular the so-called del-Pezzo surfaces) are of crucial importance \cite{Cicoli:2011it,Cicoli:2018tcq,Cicoli:2021dhg}.

In this context, there has been another divisor topology, namely the so-called ``Wilson divisor" which is necessary for realizing poly-instanton corrections \cite{Blumenhagen:2012kz}. These contributions appear as exponential corrections on top of the usual $E3$-instanton corrections leading to an schematic form of the superpotential given as $e^{- a T_s + e^{-b T_w}}$, where $a>0, b>0$ and $T_s$ corresponds to the complexified four-cycle volume wrapping the $E3$-instanton while $T_w$ corresponds to the complexified four-cycle volume of the Wilson divisor. Inclusion of these effects generate sub-leading contributions for K\"ahler moduli stabilization which can help in driving inflation as well \cite{Blumenhagen:2012ue}.

\vskip0.1cm
\noindent
{\bf (ii)~Swiss-cheese structure:}
The CY threefolds with a particular divisors topology are found to be central in realizing the so-called LARGE Volume Scenario (LVS) scheme of moduli stabilization \cite{Balasubramanian:2005zx}. These divisor topologies are the so-called ``diagonal" del-Pezzo divisors in the sense that they can be shrinked to a point-like singularity by squeezing along a single direction \cite{Cicoli:2011it,Cicoli:2018tcq,Cicoli:2021dhg}.
 
 \vskip0.1cm
\noindent
{\bf (iii)~$K3$-fibred CY threefolds:}
The CY threefolds with $K3$-fibration exhibits some peculiar properties which have interesting phenomenological implications \cite{Cicoli:2008gp,Cicoli:2011it, Cicoli:2016xae, Cicoli:2017axo}.

\vskip0.1cm
\noindent
{\bf (iv)~String-loop corrections:}
Although explicit results for string loop effects are known from the toroidal computations \cite{Berg:2005ja,Berg:2007wt}, their various insights can be extrapolated for the models based on (orientifolds of) CY threefolds as well. As argued in \cite{Berg:2007wt,Cicoli:2007xp}, the two classes of string-loop effects (known as KK-type and Winding-type) are induced with some very particular kind of brane settings. In fact, a field theoretic argument for the existence of the so-called Winding loop corrections has been already proposed in \cite{vonGersdorff:2005bf}. In addition, the KK-type correction needs the presence of non-intersecting stacks of $D7/O7$ and $O3$ planes while Winding-type effect needs intersecting stacks of $D7/O7$ configurations which intersect at some non-contractible two-cycles. However these requirements have been further revisited recently in \cite{Gao:2022uop} where it has been found that Winding-type corrections can appear more generically than what is expected from the very specific brane-setting arguments of \cite{Berg:2005ja,Berg:2007wt}, something which has been anticipated from \cite{vonGersdorff:2005bf}. 

In addition, there is a different class of string-loop effect; the so-called logarithmic loop-corrections \cite{Antoniadis:2018hqy,Antoniadis:2019rkh} for which three stacks of $D7$-branes with non-trivially specific intersection loci have been realized with the appropriate global construction in \cite{Leontaris:2022rzj}. 

\vskip0.1cm
\noindent
{\bf (v)~Higher derivative F$^4$-corrections:}
In the context of higher derivative effects, there are some terms which are beyond the two-derivative contributions via K\"ahler and superpotential, and appear directly to the scalar potential at F$^4$-order as proposed in \cite{Ciupke:2015msa}. It has been found that a topological quantity defined as $\Pi(D) = \int_{CY} c_2({\rm CY}) \wedge \hat{D}$ turns out to be of central importance where $c_2({\rm CY})$ is the second Chern-class of the CY$_3$ and $\hat{D}$ denotes the dual $(1,1)$ class corresponding to the divisor $D$ of the CY$_3$.


\subsection{Relevant Divisor Topologies}
Now we will discuss some of the divisor topologies which enter in the overall game of inducing the ``right" form of the scalar potential contributions. In fact, one may have a classification of divisor topologies into three classes:

\vskip0.1cm
\noindent
{{\bf (i)~Rigid divisors:} These divisors which we will denote as $R$ correspond to the four-cycles with unit Arithmetic genus, i.e. $\chi_{_h}(R) = 1$. For smooth and connected cases, we have
\bea
h^{0,0}(R) = 1 , \quad h^{1,0}(R) = 0 = h^{2,0}(R), \quad h^{1,1}(R) = \chi(R)-2. \nonumber
\eea
Some particular examples of such surfaces can be considered as ${\mathbb P}^2$ and the del-Pezzo surfaces dP$_n$ for $ 1 \leq n \leq 8$. Such surfaces arise by including eight generic blow-up points in ${\mathbb P}^2$, and the corresponding topological quantities are given as $\{\chi_{_h} (dP_n) =1, \, \chi(dP_n) = n+3\}$ where $n = 0$ corresponds to ${\mathbb P}^2$. Because of satisfying the Witten's unit Arithmetic genus condition \cite{Witten:1996bn} which is necessary for contributing to the non-perturbative superpotential, such  divisors have been of great interest in phenomenological model building and have attracted tremendous amount of interests in recent years, e.g. \cite{Blumenhagen:2010ja,Cicoli:2011it,Cicoli:2011qg,Gao:2013pra,Cicoli:2013cha} for initial attempts of concrete global model building.
}

\vskip0.1cm
\noindent
{{\bf (ii)~Non-rigid divisors:} These divisors which we will denote as $K$ can be described as divisors with non-zero deformations in the CY threefold, i.e. $h^{2,0}(K) \neq 0$, and $h^{1,0}(K) = 0$ which leads to $\chi_{_h}(K) > 1$. In fact one has the following Hodge numbers to characterise such topologies,
\bea
h^{0,0}(K) = 1 , \quad h^{1,0}(K) = 0 , \quad h^{2,0}(K) = \chi_{_h}-1, \quad h^{1,1}(K) =\chi - 2 \chi_{_h}. \nonumber
\eea
The $K3$  surfaces are examples with $\chi_{_h}(K3) = 2$ and $\chi(K3) = 24$. Another example is the so-called `special deformation' divisor (${\rm SD}$) which appear very often in the overall divisor topology scan. One class of such ${\rm SD}$ divisor is described by $\chi_{_h}({\rm SD}) = 3, \, \chi({\rm SD}) = 36$. We note that such a divisor can be ``rigidified" by turning-on magnetic fluxes to contribute to the non-perturbative superpotential \cite{Bianchi:2011qh, Louis:2012nb}.
}

\vskip0.1cm
\noindent
{{\bf (iii)~Wilson divisors:} These divisors  denoted as $W$ can be described as rigid but non-simply connected divisors having zero deformations in the CY threefold, i.e. $h^{2,0}(W) = 0$ but $h^{1,0}(W) \neq 0$ leading to $\chi_{_h}(W) \leq 0$ for $W$ being smooth. Hodge numbers to characterize such topologies are,
\bea
& & \hskip-0.75cm h^{0,0}(W) = 1 , \quad h^{1,0}(W) = 1 - \chi_{_h}, \nonumber\\
& & \hskip-0.75cm h^{2,0}(W) = 0, \quad h^{1,1}(W) =\chi + 2 - 4 \chi_{_h}. \nonumber
\eea
An explicit example is the so-called Wilson surface used for poly-instanton effects corresponding to $\chi_{_h}(W) = 0$ and $\chi(W) = 0$ \cite{Blumenhagen:2012kz}. Having both $\chi$ and $\chi_{_h}$ zero makes this surface quite peculiar \cite{Blumenhagen:2012kz}.
}

\vskip0.1cm
\noindent
{{\bf (iv)~Diagonal divisors:} A divisor $D_i$ if diagonal if it satisfies \cite{Cicoli:2011it, Cicoli:2018tcq},
\bea
\label{eq:diag-divisor}
& & \hskip-1cm D_{\rm diag}: \quad \kappa_{iii} \, \, \kappa_{i j k } = \kappa_{ii j} \, \, \kappa_{ii k}\,; \, \quad \quad i \,\,\, {\rm fixed} \quad {\rm and} \quad \forall \, \, \, j, k.
\eea
If this diagonality condition is fulfilled for divisors with non-vanishing cubic self-intersections i.e. $\kappa_{iii} \neq 0$, then there exists a basis of coordinate divisors such that the volume of each of the four-cycles $D_i$ ($\tau_i$) is a complete-square quantity in terms of the two-cycle volume moduli ($t^i$):
\bea
\label{eq:kappa0}
& & \hskip-1cm \tau_i \equiv \frac{1}{2}\, \kappa_{ijk} t^j \, t^k = \frac{1}{2 \, \kappa_{iii}}\, \kappa_{ii j} \, \kappa_{ii k} t^j \, t^k = \frac{1}{2 \, \kappa_{iii}}\, \left(\kappa_{iij} \,t^j \, \right)^2.
\eea
}

\vskip0.1cm
\noindent
{{\bf (v)~(Diagonal) del-Pezzo divisors:} The popular del-Pezzo surfaces - giving the Swiss-Cheese structure of the CY threefold - are defined as projective algebraic surfaces with ample anti-canonical divisor class $(-{\cal K})$ so that $-{\cal K} \cdot {\cal C} > 0$ for each curve ${\cal C}$. The del-Pezzo divisors $D_s$ satisfy the following two necessary conditions \cite{Cicoli:2011it},
\bea
\label{eq:dP}
& & \int_{X} \hat{D}_s \wedge \hat{D}_s \wedge \hat{D}_s = \kappa_{sss} > 0\, , \qquad \int_{X} \hat{D}_s \wedge \hat{D}_s \wedge \hat{D}_i \leq 0 \qquad \forall \, i \neq s \,.
\eea
Here the self-triple-intersection number $\kappa_{sss}$ corresponds to the degree of the del-Pezzo four-cycle dP$_n$ where $\kappa_{sss} = 9 - n$, which is always positive as $n \leq 8$ for del-Pezzo surfaces. We note that dP$_0$ is ${\mathbb P}^2$ with $\kappa_{sss} = 9$. 
Among the Fano surfaces we find ${\mathbb P}^1 \times {\mathbb P}^1$ (referred as the Hirzebruch surface ${\mathbb F}_0$ as  well), with $h^{1,1} = 2$.

\vskip0.1cm
\noindent
{{\bf (vi)~Divisors of vanishing second Chern number:} These are relevant for higher derivative F$^4$-corrections as we have discussed earlier. It turns out that $\Pi(D) = 2 \, \chi(D) - 12 \, \chi_h(D)$, and subsequently the condition for divisors of vanishing $\Pi$ boils down to 
\bea
& &  \hskip-0.5cm \chi(D) = 6 \, \chi_h(D) \quad \Leftrightarrow \quad h^{1,1}(D) = 4 \, h^{0,0}(D) - 2\, h^{1,0}(D) + 4 \, h^{2,0}(D). \nonumber
\eea
So, the surfaces such as dP$_3$, ${\mathbb T}^4$, and Wilson-type with $h^{1,1}(W) = 2$ indeed satisfy this property and fall in the category of vanishing $\Pi$ divisors.

}

\section{Classifying Divisor Topologies For Phenomenology}
\label{sec:Topology}
In this section, we analyze the divisor topologies corresponding to a subset of the CY threefold geometries arising from the four-dimensional reflexive polytopes of the Kreuzer-Skarke database \cite{Kreuzer:2000xy}, which motivates for some pheno-inspired classification. To begin with, we consider the following input data for constructing explicit global models:
\begin{itemize}
\item GLSM charges
\item Stanley-Reisner (SR) ideal
\item Second Chern-class $c_2(CY)$ of the CY threefold
\item Intersection tensor $\kappa_{ijk}$
\item Fundamental group
\end{itemize}
Subsequently, using these CY data and some packages such as \texttt{cohomCalg} package  \cite{Blumenhagen:2010pv,Blumenhagen:2011xn}, and \href{https://cytools.liammcallistergroup.com/about/}{\texttt{CYTools}}, we investigate the various divisor (and curve) topologies to seek for their relevance for generating the scalar potential pieces in a given model as discussed in the previous section.

\subsection{Methodology}
We focus on looking at the topology of the toric divisors, also called as ``coordinate divisors" $D_i$, which are defined through setting the toric coordinates to zero, i.e. $x_i = 0$. Though such a choice can heavily limit the number of divisors which can have deformations inside the CY threefold, this strategy should still be sufficient for capturing the rigid four-cycles such as the del-Pezzo surfaces relevant for LVS model building \cite{Balasubramanian:2005zx} and `Wilson' divisors suitable for poly-instanton effects \cite{Blumenhagen:2012kz}. Classifying divisor topologies limited to the toric divisors has been motivated earlier for scanning the del-Pezzo divisors in the LVS model building, e.g. see \cite{Cicoli:2018tcq,Cicoli:2021dhg}. Further, we consider the so-called ``favorable" triangulations and favorable CY geometries in the sense that all the toric divisors of the CY threefolds descend from the Ambient fourfold. Moreover, following the strategy of \cite{Shukla:2022dhz}, we also exclude a couple of additional examples which have non-trivial fundamental groups and may not have conventional phenomenological interests due to non-smooth internal structures at divisor level. 

For a generic divisor ($D$) of the CY threefold $X$, there are only four independent Hodge numbers, namely $h^{0,0}, h^{1,0}, h^{2,0}$ and $h^{1,1}$. Two of these can be computed from the Euler characteristics $\chi(D)$ and the Arithmetic genus $\chi_{_h}(D)$  via knowing the second Chern class of the CY $(c_2(X))$ along with the classical triple intersection numbers $\kappa_{ijk}$, (e.g. see \cite{Blumenhagen:2008zz,Collinucci:2008sq, Bobkov:2010rf,Cicoli:2016xae}),
\bea
& & \hskip-0.4cm \chi(D) = 2 h^{0,0} - 4 h^{1,0} + 2 h^{2,0} + h^{1,1}= \int_{X} \left(\hat{D} \wedge \hat{D} \wedge \hat{D} + c_2(X) \wedge \hat{D} \right), \, \nonumber\\
& & \hskip-0.4cm \chi_{_h}(D) = h^{0,0} - h^{1,0} + h^{2,0} = \frac{1}{12} \int_{X}\left(2\, \hat{D} \wedge \hat{D} \wedge \hat{D} + c_2(X) \wedge \hat{D} \right),
\label{eq:chi-chih}
\eea
where $\hat{D}$ denotes the 2-forms dual to the divisor class.  Using \texttt{HodgeDiamond} and \texttt{Lambda0CotangentBundle} modules of  \texttt{cohomCalg} package \cite{Blumenhagen:2010pv,Blumenhagen:2011xn}, or \href{https://cytools.liammcallistergroup.com/about/}{\texttt{CYTools}}, the Hodge numbers of all the divisors can be easily computed.


\subsection{Scanning Results}
Now we present the scanning results for the divisor topologies corresponding to the favorable CY geometries arising from both the CY datasets, namely AGHJN database \cite{Altman:2014bfa} (or THCYs) and the pCICY database \cite{}. 

\noindent
{\bf (i)~Divisor topologies for THCYs: } For AGHJN database we consider the favorable CYs with a trivial fundamental group resulting in {\bf 139740} toric divisors corresponding to 15829 CYs listed in Table \ref{tab_number-of-space-and-divisors}. 

\begin{table}[H]
\centering
\resizebox{\textwidth}{!}{%
\begin{tabular}{|c||c|c|c|c||c|c|} 
\hline
$h^{1,1}$ & polytope & Triang & Geom & fav-Geom & fav-Geom$^\ast$ & divisors of fav-Geom$^\ast$ \\
 \hline
 \hline
 1 & 5 & 5 & 5 & 5 & 4 & 20  \\
 2 & 36 & 48  & 39  & 39 & 37 & 222 \\
 3 & 244 & 569 & 306 & 305 & 300 & 2100 \\
 4 & 1197 & 5398 & 2014 & 2000 & 1994 & 15952 \\
 5 & 4990 & 57132 & 13635 & 13494 & 13494 & 121446 \\
\hline
Tot \# & 6472 & 63152 & 15999 & 15843 & 15829 & {\bf 139740} \\ 
 \hline
\end{tabular}
}
\caption{Number of CYs and their toric divisors for $1\leq h^{1,1}(CY) \leq 5$ \cite{Shukla:2022dhz}. }
\label{tab_number-of-space-and-divisors}
\end{table}

\begin{table}[H]
\centering
\resizebox{\textwidth}{!}{%
\begin{tabular}{|c||c|c|c|c|} 
\hline
Type & Divisor topology   & Distinct topology & Frequency & {$h^{1,1}$(CY)} \\
& $\{h^{0,0}, h^{1,0}, h^{2,0}, h^{1,1}\}$ & (out of 565) & (out of 139740) &  \\
\hline
$R_n$ & $\{1, 0, 0, n\}$  & 63 & 76839 & 2-5  \\
$K_n^m$ & $\{1, 0, m, n\}$  & 395 & 55972 & 1-5  \\
$W_n^m$ & $\{1, m, 0, n\}$  & 107 & 6929 & 2-5 \\
\hline
\end{tabular}
}
\caption{Divisor topologies and their frequencies with $m, n \in {\mathbb Z}_+$ \cite{Shukla:2022dhz}.}
\label{tab_divisor-topologies-THCYs}
\end{table}

\begin{table}[H]
	\centering
	\resizebox{\textwidth}{!}{%
	\hskip0.11cm \begin{tabular}{|c||c|c||c|c|c|c|c|c||c|}
		\hline
		$h^{1,1}$ & Poly$^*$ &  Geom$^*$ & ${\mathbb P}^2$  & ${\mathbb P}^1 \times {\mathbb P}^1$  & $\mathrm{ddP}_n$ & $\mathrm{ddP}_6$  & $\mathrm{ddP}_7$  & $\mathrm{ddP}_8$  & $n_{\rm LVS}$ \\
		&  & ($n_{\rm CY}$) &  &  & $1\leq n \leq 5$ &  &  &  & ($\mathrm{ddP}_n\geq 1$) \\
		\hline
		1 & 5 & 5 & 0 & 0  & 0  & 0 & 0 & 0 & 0 \\
		2 & 36 & 39 & 9 & 2 & 0 & 2  & 4  & 5 & 22 \\
		3 & 243 & 305 & 59 & 16 & 0  & 17 & 40 & 39 & 132\\
		4 & 1185 & 2000 & 372 & 144 & 0 & 109 & 277 & 157 &  750 \\
		5 & 4897 & 13494 & 2410 & 944  & 0  & 624 & 827 & 407 & 4104 \\
		\hline
	\end{tabular}
	}
	\caption{CY geometries with a ddP divisor suitable for LVS \cite{Cicoli:2021dhg,Shukla:2022dhz}.}
	\label{tab_ddPns-GstarM}
\end{table}

\begin{table}[H]
  \centering
  \resizebox{\textwidth}{!}{%
 \begin{tabular}{|c||c|c||c|c|c|c||c|c|c|c|}
\hline
$h^{1,1}$ & Poly$^*$  & Geom$^*$  & At least  & Single & Two & Three  & $n_{\rm LVS}$ \&  & $n_{\rm LVS}$ \& & $n_{\rm LVS}$ \&     \\
&  & $(n_{CY})$ & one $W_\Pi$ & $W_\Pi$ & $W_\Pi$  & $W_\Pi$  & 1 $W_\Pi$  & 2 $W_\Pi$  & 3 $W_\Pi$     \\
 \hline
 1 & 5 & 5 & 0 & 0  & 0 & 0 & 0  & 0 & 0   \\
 2 & 36 & 39 & 0 & 0 & 0 & 0 & 0  & 0 &  0 \\
 3 & 243 & 305 & 19 & 19 & 0 & 0 & 4 & 0 &  0  \\
 4 & 1185 & 2000 & 210 & 202 & 8 & 0 & 62  & 1 &  0   \\
 5 & 4897 &13494 & 1764  & 1599  & 154 & 11 &  442 & 79  &  1 \\
 \hline
  \end{tabular}
  }
  \caption{CYs with vanishing $\Pi$ divisors $W_\Pi$, and a ddP$_n$ for LVS \cite{Shukla:2022dhz,Cicoli:2023njy}.}
   \label{tab_GwilsonPi0LVS}
 \end{table}

\noindent
The rigid, non-rigid and Wilson-type divisor topologies are listed in Table \ref{tab_divisor-topologies-THCYs} while diagonal del-Pezzo divisors relevant for LVS model building are listed in Table \ref{tab_ddPns-GstarM}. Further, the divisors of vanishing $\Pi$ relevant for bypassing higher derivative F$^4$-corrections are listed in Table \ref{tab_GwilsonPi0LVS}.

\vskip0.1cm
\noindent
{\bf (ii)~Divisor topologies for pCICYs:} A detailed classification of all the toric divisor topologies for the pCICYs database \cite{Anderson:2017aux} have been classified in \cite{Carta:2022web}. Considering the favorable examples only, there are a total of 57885 toric divisors with non-identical GLSM charges for a total of 7868 pCICY favorable geometries. Quite surprizing to us, after analyzing all the 57885 toric divisors we have found that there are only 11 types of divisor topologies as listed in Table \ref{tab_divisor-topologies}. 
\begin{table}[H]
\centering
\resizebox{\textwidth}{!}{%
\begin{tabular}{|c||c|c|c|c|c|} 
\hline
{Sr.\#}  & Divisor  & frequency & frequency & $h^{1,1}$ & $\displaystyle{\int_{_{\rm CY}} \hat{D}^3}$ \\
  & topology & (57885 divisors) & (7820 spaces) & (pCICY) &  \\
\hline
T1 & $\{1, 0, 1, 20\}$ & 30901 & 7736 & 2-15 & 0 \\
T2 & $\{1, 0, 2, 30\}$ & 22150 & 7436 & 2-15 & 0 \\
T3 & $\{1, 0, 3, 38\}$ & 3372 & 2955 & 2-13 & 2 \\
T4 & $\{1, 0, 3, 36\}$ & 91 & 91 & 3-13 & 4 \\
T5 & $\{1, 0, 4, 46\}$ & 714 & 690 & 2-11 & 4 \\
T6 & $\{1, 0, 4, 45\}$ & 283 & 277 & 1-11 & 5 \\
T7 & $\{1, 0, 4, 44\}$ & 91 & 91 & 2-11 & 6 \\
T8 & $\{1, 0, 5, 52\}$ & 198 & 198 & 1-9 & 8 \\
T9 & $\{1, 0, 5, 51\}$ & 28 & 28 &  1-9 & 9 \\
T10 & $\{1, 0, 6, 58\}$ & 42 & 42 & 1-7 & 12 \\
T11 & $\{1, 0, 7, 64\}$ & 15 & 15 & 1-5 & 16 \\
\hline
\end{tabular}
}
\caption{Divisor topologies with $\{h^{0,0}, h^{1,0}, h^{2,0}, h^{1,1}\}$ for favorable pCICYs and their frequencies of appearance in the scan \cite{Carta:2022web}.}
\label{tab_divisor-topologies}
\end{table}
\noindent
Some interesting observations for pCICY divisor topologies are:
\begin{itemize}
\item The first two topologies T1 and T2, which corresponds to the K3 surfaces and the so-called special deformation (SD) divisors appear most frequently. In fact, more than 53000 out of a total of 57885 divisors fall in these two categories ! 
\item We observe that the self triple intersection number vanishes for divisor topologies T1 and T2 which satisfy $h^{1,1}(D) = 10 + 10 \, h^{2,0}(D)$. 
\item We find that there are neither rigid surfaces not any ${\mathbb T}^4$ surfaces present in the entire toric divisor topologies for favorable pCICYs. 
\end{itemize}
\noindent

\subsection{LVS Inflationary Models}
\label{sec:LVS-inflation}
After fixing the overall volume ${\cal V}$ of the CY threefold in the LVS scheme, there still remain several unfixed moduli, which are usually referred as `LVS flat directions'. Being protected by no-scale symmetry at leading orders, these leftover moduli are very good candidates for driving inflation if a nearly flat potential is ensured for them via the sub-leading effects. Depending on the geometries of these moduli and the types of corrections to the inflationary potential one can divide them in three main types, namely Blow-up inflation, Fibre inflation and Poly-instanton inflation. Scanning results for CY$_3$s with the minimal global model requirement for these models are as follows:

\vskip0.1cm
\noindent
{\bf (i)~Blow-up inflation \cite{Conlon:2005ki,Blanco-Pillado:2009dmu,Cicoli:2017shd}:} It needs two diagonal dP divisors; one for realizing LVS and another for driving inflation. Scanning results for suitable CYs are listed in Table \ref{tab_blowup-Gstar}.
\begin{table}[H]
\centering
\resizebox{\textwidth}{!}{%
\hskip0.11cm \begin{tabular}{|c|c|c||c|c|c|c||c|c|}
\hline
 $h^{1,1}$ & Poly$^*$ &  Geom$^*$ & $n_{\rm ddP}=1$  & $n_{\rm ddP}=2$ & $n_{\rm ddP}=3$  & $n_{\rm ddP}=4$ & $n_{\rm LVS}$ & Blow-up \\
&  & $(n_{\rm CY})$ &  &   &  &  &  & infl. \\
 \hline
 1 & 5 & 5 & 0 & 0  & 0  & 0 & 0 & 0 \\
 2 & 36 & 39 & 22 & 0  & 0  & 0 & 22 & 0 \\
 3 & 243 & 305 & 93 & 39  & 0 & 0 & 132 & 39 \\
 4 & 1185 & 2000 & 465 & 261 & 24 & 0 & 750 & 285 \\
 5 & 4897 & 13494 & 3128  & 857  & 106 & 13 & 4104 & 976 \\
 \hline
  \end{tabular}
  }
\caption{Number of LVS CY geometries suitable for blow-up inflation \cite{Cicoli:2023njy}.}
   \label{tab_blowup-Gstar}
 \end{table}

\vskip-0.1cm
\noindent
{\bf (ii)~Fibre inflation \cite{Cicoli:2008gp,Cicoli:2016chb,Cicoli:2016xae,Cicoli:2017axo, Cicoli:2024bxw}:} It needs a K3- or ${\mathbb T}^4$ fibred CY threefolds with at least one diagonal dP divisor to have LVS. The local brane setting should be suitable for having the appropriate string-loop corrections of KK and Winding type. Scanning results for suitable CYs are listed in Table \ref{tab_fibre-Gstar} which are consistent with the previous subsets of scans reported in \cite{Cicoli:2016xae, Cicoli:2011it}. 
\begin{table}[H]
 \centering
 \resizebox{\textwidth}{!}{%
\hskip0.11cm \begin{tabular}{|c|c|c||c|c|c||c|}
\hline
 $h^{1,1}$ & Poly$^*$ &  Geom$^*$ & $n_{\rm LVS}$ & K3 fibred  & $n_{\rm LVS}$ with K3 fib.& $n_{\rm LVS}$ with \\
   &  & $(n_{\rm CY})$  & &  CY & (fibre inflation)  &  K3 fib. \& $D_\Pi$ \\
 \hline
 1 & 5 & 5 & 0 & 0  & 0  &  0  \\
 2 & 36 & 39 & 22 & 10  & 0  &  0  \\
 3 & 243 & 305 & 132 & 136  & 43 \cite{Cicoli:2016xae}  &  0  \\
 4 & 1185 & 2000 & 750 & 865 & 171  & 28  \\
 5 & 4897 & 13494 & 4104  & 5970  &  951 & 179 \\
 \hline
  \end{tabular}
 }
  \caption{Number of LVS CY geometries suitable for fibre inflation \cite{Cicoli:2023njy}.}
    \label{tab_fibre-Gstar}
 \end{table}

\vskip-0.1cm
\noindent
{\bf (iii)~Poly-instanton inflation \cite{Cicoli:2011ct,Blumenhagen:2012ue,Gao:2013hn,Gao:2014fva}:} It needs one Wilson divisor of type $h^{1,0}(W) =1 = h^{1,0}_+(W)$ for a suitable holomorphic involution. The scanning results for suitable CY threefolds are listed in Table \ref{tab_GwilsonLVS}.

\begin{table}[H]
  \centering
  \resizebox{\textwidth}{!}{%
 \begin{tabular}{|c||c|c||c|c|c||c|c|c|c|}
\hline
$h^{1,1}$ & Poly$^*$  & Geom$^*$  & Single & Two & Three  & $n_{\rm LVS}$ & $n_{\rm LVS}$ \& $W$ & $n_{\rm LVS}$ \&  $W_\Pi$   \\
&  & $(n_{\rm CY})$ & $W$ & $W$ & $W$  &  & (poly-inst.) & (topol. tamed)     \\
 \hline
 1 & 5 & 5 & 0 & 0  & 0 & 0 & 0  & 0    \\
 2 & 36 & 39 & 0 & 0 & 0 & 22 & 0  & 0  \\
 3 & 243 & 305 & 19 & 0 & 0 & 132 & 4 & 4   \\
 4 & 1185 & 2000 & 221 & 8 & 0 & 750 & 75  & 63   \\
 5 & 4897 &13494 & 1874  & 217  & 43 & 4104 &  660 & 522  \\
 \hline
  \end{tabular}
  }
\caption{Number of LVS CY geometries suitable for poly-instanton inflation. Here $W$ denotes a generic Wilson divisors, while $W_\Pi$ has $\Pi=0$ \cite{Cicoli:2023njy}.}
\label{tab_GwilsonLVS}
\end{table}
\noindent
In fact, it has been suggested recently that one cannot avoid the presence of string-loop correction in a Blow-up inflation setup leading to a new class of inflationary proposal called as Loop Blow-up inflation \cite{Bansal:2024uzr}. Also, some alternate realizations of fibre inflation in perturbative LVS with non Swiss-Cheese CYs have been recently proposed in \cite{Bera:2024ihl,Leontaris:2025hly,Leontaris:2025xit,Leontaris:2026sqh}.

Finally, let us stress that in all our scans we have only focused on the minimal requirements to realise explicit global constructions of LVS inflationary models. However, every model has to be engineered in a specific way on top of fulfilling the first order topological requirements. 

\section{Multi-field Approach to Fibre Inflation}
\label{sec:inflaton-bound}

\subsection{Recent Challenges For Single-Field Fibre Inflation}
In the minimal Fibre inflation model based on CY$_3$ with $h^{1,1} = 3$, two moduli, namely ${\cal V}$ and $\tau_s$, are fixed by the standard LVS while the third one corresponds to the inflaton field $\tau_f$. The crucial thing to note here is the fact that although $\tau_f$ remains flat after fixing the overall volume ${\cal V}$, there exists a bound on the field range it can traverse during inflationary dynamics. This arises because of the fact that the K\"ahler cone conditions (KCC) $\int_{C_i} J > 0$ hold for all the curves $C_i$ in the Mori cone of the CY$_3$ which typically translates into the following generic constraint:
\bea
\label{eq:KC-gen}
& & \hskip-1cm {\rm KCC:} \qquad m_{\alpha\beta}\,t^\beta > 0, \quad {\rm for \, \, some} \quad m_{\alpha\beta} \in {\mathbb Z},
\eea
where summation for $\beta$ runs in $h^{1,1}({\rm CY})$ while $\alpha$ corresponds to the number of K\"ahler cone conditions which is $h^{1,1}({\rm CY})$ for simplicial cases but can be more for non-simplicial cases. Therefore depending on the values of $m_{\alpha\beta}$ in a given concrete model, there will be constraints on the 2-cycle volume and subsequently on the 4-cycle volumes. For example, taking the volume-form ${\cal V} = \frac12 k_{bbf} (t^b)^2 t^f + \frac16 k_{sss} (t^s)^3$ and using $\tau_b = k_{bbf} t^b t^f$, $\tau_f = \frac12{k_{bbf}}(t^b)^2$, $\tau_s = \frac12{k_{sss}} (t^s)^2$, a generic KCC of the form $m_s t^s + m_b t^b + m_f t^f > 0$ translates into,
\bea
\label{eq:KC-gen}
& & \hskip-1.3cm \frac{m_f}{\tau_f} \left({\cal V} + \lambda_s \tau_s^{3/2} \right) + 2\sqrt2\, \lambda_b \,m_b \sqrt{\tau_f} > 3 \lambda_s\, m_s \sqrt{\tau_s}.
\eea
Such a situation typically results in two classes of upper bounds in explicit global models \cite{Cicoli:2017axo, Cicoli:2018tcq}. First one corresponds to $m_s = 0$ leading to $\tau_f < {\cal V}^{2/3}$ while one gets a better case $\tau_f < {\cal V}/\sqrt{\tau_s}$ for $m_b = 0$. For cases with $m_f = 0$, one typically gets a lower bound for $\tau_s$ such that $\tau_s < c_1 \tau_f$ for $c_1 >0$. It turns out that the field-range-bound problem is rooted in the presence of rigid dP divisor and the Swiss-Cheese structure of the CY, and alternate proposal of realizing fibre inflation in perturbative LVS have been proposed recently \cite{Bera:2024ihl,Leontaris:2025hly,Leontaris:2025xit}.

\subsection{Assisted Fibre Inflation}

{\bf Global model:} We consider a CY$_3$ with $h^{1,1} = 3$ corresponding to the polytope Id: 249 in the CY database of \cite{Altman:2014bfa} which has been studied earlier in \cite{Gao:2013pra, Leontaris:2022rzj, Bera:2024zsk}. It is described by the following toric data:
\begin{center}
\begin{tabular}{|c|ccccccc|}
\hline
Hyp &  $x_1$  & $x_2$  & $x_3$  & $x_4$  & $x_5$ & $x_6$  & $x_7$       \\
\hline
4 & 0  & 0 & 1 & 1 & 0 & 0  & 2   \\
4 & 0  & 1 & 0 & 0 & 1 & 0  & 2   \\
4 & 1  & 0 & 0 & 0 & 0 & 1  & 2   \\
\hline
& $K3$  & $K3$ & $K3$ &  $K3$ & $K3$ & $K3$  &  SD  \\
\hline
\end{tabular}
\end{center}
The Hodge numbers are $(h^{2,1}, h^{1,1}) = (115, 3)$, the Euler number is $\chi=-224$ while the Stanley-Reisner ideal is ${\rm SR} =  \{x_1 x_6, \, x_2 x_5, \, x_3 x_4 x_7 \}$. The analysis of the divisor topologies using {\it cohomCalg} \cite{Blumenhagen:2010pv, Blumenhagen:2011xn} shows that the first six toric divisors are K3 surfaces while the seventh one is described by Hodge numbers $\{h^{0,0} = 1, h^{1,0} = 0, h^{2,0} = 27, h^{1,1} = 184\}$. Considering the divisor basis $\{D_1, D_2, D_3\}$ and the K\"ahler form $J = t^1 D_1+t^2 D_2+t^3 D_3$, the second Chern class $c_2({\rm CY})$ is given as $c_2({\rm CY}) = 5 D_3^2+12 D_1 D_2 + 12 D_2 D_3+12 D_1 D_3$, while $k_{123} = 2$ is the only non-zero intersection leading to ${\cal V} = 2 \, t^1\, t^2\, t^3 = \frac{1}{\sqrt{2}}\,\sqrt{\tau_1 \, \tau_2\, \tau_3}$ where $\tau_1 = 2 t^2 t^3, \tau_2 = 2 t^1 t^3, \tau_3 = 2 t^1 t^2$ and ${\cal V} = t^1 \tau_1 = t^2 \tau_2  = t^3 \tau_3$ as in toroidal case. The K\"ahler cone conditions are:
\bea
\label{eq:KC-torus}
& & \hskip-1.5cm \text{KCC:} \qquad  t^1 > 0, \quad t^2 > 0, \quad t^3 > 0.
\eea
Note that unlike KCC in Eq.~(\ref{eq:KC-gen}) corresponding to a CY$_3$ with a dP divisor \cite{Cicoli:2018tcq}, the KCC in Eq.~(\ref{eq:KC-torus}) do not put any other bound except $\tau_\alpha > 0$ and ${\cal V} > 0$. Therefore, the $\tau_3$ modulus is not heavily constrained if two of them are already fixed by some moduli stabilisation scheme.

\vskip0.2cm
\noindent
{\bf Choice of Involution and Brane setting:} For a given holomorphic involution, one needs to introduce D3/D7-branes and fluxes in order to cancel all the D3/D7 tadpoles. In fact, one can nullify the D7-tadpoles via introducing stacks of $N_a$ D7-branes wrapped around suitable divisors (say $D_a$) and their images ($D_a^\prime$). However, the presence of D7-branes and O7-planes also contributes to the D3-tadpoles, which, in addition, receive contributions from  $H_3/F_3$ fluxes, D3-branes and O3-planes. One needs to satisfy  \cite{Blumenhagen:2008zz}:
\bea
\label{eq:D3D7tadpole}
& & \hskip-0.5cm {\bf \rm D7:} \,\, \sum_a\, N_a \left([D_a] + [D_a^\prime] \right) = 8\, [{\rm O7}],\\
& & \hskip-0.5cm {\bf \rm D3:} \,\, N_3 = \frac{N_{\rm O3}}{4} + \frac{\chi({\rm O7})}{12} + \sum_a\, \frac{N_a \left(\chi(D_a) + \chi(D_a^\prime) \right) }{48}.\nonumber
\eea
where $N_3 \equiv N_{\rm D3} + \frac{N_{\rm flux}}{2} + N_{\rm gauge}$ such that $N_{\rm D3}$ is the net number of D3-brane, $N_{\rm flux} = (2\pi)^{-4} (\alpha^\prime)^{-2}\int_X H_3 \wedge F_3$ is the contribution from background fluxes and $N_{\rm gauge} = -\sum_a (8 \pi)^{-2} \int_{D_a}\, {\rm tr}\, {\cal F}_a^2$ is due to D7 worldvolume fluxes. Considering the involution $x_7 \to - x_7$ results in fixed point set $\{O7 = D_7\}$ without any $O3$-planes, and subsequently one can consider 3 stacks of $D7$-branes wrapping each of the three divisors $\{D_1, D_2, D_3\}$, and the D3/D7-brane tadpoles are nullified via $8\, [O_7] = 8 \left([D_1] + [D_1^\prime] \right) + 8 \left([D_2] + [D_2^\prime] \right)+ 8 \left([D_3] + [D_3^\prime] \right)$ resulting in $N_3 = 44$.

\vskip0.2cm
\noindent
{\bf Scalar potential:} The scalar potential receives contributions from BBHL's $(\alpha^\prime)^3$ corrections, string loop corrections of `log-loop' type and Winding-type while KK-type corrections are absent. In addition, there are higher derivative F$^4$-corrections appearing at order $(\alpha^\prime)^3$ itself. The three-field potential is given as:
\bea
\label{eq:Vfinal-simp3}
& & \hskip-0.2cm V \equiv V({\cal V}, t^2, t^3) = \frac{{\cal C}_{\rm up}}{{\cal V}^p} +  \frac{{\cal C}_1}{{\cal V}^3} \left(\hat\xi + 2\,\hat\eta \, \ln{\cal V} - 8\,\hat\eta + 2\,\hat\sigma \right) \\
& & \hskip-0.2cm - \frac{{\cal C}_2}{{\cal V}^3} \biggl(2\,{\cal C}_{w_1} \frac{t^2 \,t^3}{{\cal V}} +  \frac{{\cal C}_{w_2}}{t^2} + \frac{{\cal C}_{w_3}}{t^3} + \frac{{\cal C}_{w_4}\,t^2 t^3}{{\cal V} + 2 (t^2)^2 t^3} +  \frac{{\cal C}_{w_5}}{2(t^2+t^3)} + \frac{{\cal C}_{w_6}\,t^2 t^3}{{\cal V} + 2 t^2 (t^3)^2} \biggr)\nonumber\\
& &  \hskip-0.2cm  \, +  \frac{{\cal C}_3}{{\cal V}^3}\,\left(\frac{1}{2 \, t^2 \, t^3} + \frac{t^2}{\cal V} + \frac{t^3}{\cal V} \right) + \cdots,\nonumber
\eea
where $\mathcal{C}_{w_\alpha}$ are complex-structure moduli dependent quantities, and ${\cal C}_{\rm up}$ is the uplifting term needed to achieve a de Sitter vacuum with $p = 4/3$ corresponding to anti-D3 uplifting \cite{Kachru:2003aw,Crino:2020qwk,Cicoli:2017axo,AbdusSalam:2022krp}, and $p = 2$ for D-term uplifting \cite{Burgess:2003ic,Achucarro:2006zf,Braun:2015pza} while  $p = 8/3$ for the T-brane uplifting \cite{Cicoli:2015ylx,Cicoli:2017shd}.  In addition, the various coefficients ${\cal C}_i$'s are given by,
\bea
\label{eq:calCis}
& & \hskip-0.5cm {\cal C}_1 = \frac{3\kappa |W_0|^2}{4} = \frac{3{\cal C}_2}{4}, \,{\cal C}_3 = - \frac{24\, \lambda\,\kappa^2\, |W_0|^4}{g_s^{3/2}}, \, |\lambda| =  \, {\cal O}(10^{-4}), \, \kappa = \frac{g_s\, e^{K_{cs}}}{8\pi}, \nonumber\\
& & \hskip-0.5cm \hat\xi = \frac{\xi}{g_s^{3/2}}~, \quad \hat\sigma = g_s^{1/2}\, \sigma~,\quad \hat\eta = g_s^{1/2}\, \eta~, \quad \frac{\hat\xi}{\hat\eta} = -\frac{\zeta[3]}{\zeta[2]\,g_s^2\, \eta_0}, \quad \frac{\hat\sigma}{\hat\eta} = -\frac{\sigma_0}{\eta_0}. \nonumber
\eea

\vskip0.2cm
\noindent
{\bf Multi-field Inflationary Dynamics:} The multi-field dynamics are governed by the equations of motion using the number of e-folds $N$ as the time coordinate ($dN = H dt$):
\begin{align}
\label{eq:FieldEquation1}
\frac{d^2\Phi^a}{dN^2} &+ \Gamma^a_{bc}\frac{d\Phi^b}{dN}\frac{d\Phi^c}{dN} + \left(3 + \frac{1}{H}\frac{dH}{dN}\right)\frac{d\Phi^a}{dN} + \frac{\mathcal{G}^{ab}\partial_b V}{H^2} = 0,
\end{align}
with the Friedmann equation:
\begin{align}
\label{eq:FieldEquation2}
H^2 = \frac{1}{3}\left(V(\Phi^a) + \frac{1}{2}H^2\mathcal{G}_{ab}\frac{d\Phi^a}{dN}\frac{d\Phi^b}{dN}\right).
\end{align}
Starting with a scalar potential given as a function of the moduli fields $\Phi^a$, the following form of the field equations (\ref{eq:FieldEquation1}) can be directly useful,
\bea
\label{eq:EOM2}
& & \hskip-0.95cm \frac{d^2\Phi^a}{dN^2}+{\Gamma^a}_{bc} \frac{d\Phi^b}{dN} \frac{d\Phi^c}{dN}+\left(3- \frac{1}{2} \, {\cal G}_{ab} \frac{d\Phi^a}{dN} \frac{d\Phi^b}{dN} \right) \left(\frac{d\Phi^a}{dN}+ \frac{{\cal G}^{ab} \partial_b V}{V} \right)=0,
\eea
where the scalar potential $V$, the metric ${\cal G}_{ab}$ and ${\cal G}^{ab}$ are explicit functions of the fields $\Phi^a$ and can be estimated through the tree level K\"ahler potential. 

Subsequently, once the effective potential is determined in a given concrete global multi-field model, one can numerically solve the above second-order differential equations (\ref{eq:EOM2}), and get the evolution of the field trajectories $\Phi^a(N)$ in terms of the number of $e$-folds. Subsequently, the cosmological observables such as the scalar power spectrum $P_s$, the spectral index $n_s$, its running $\alpha_s$, and the tensor-to-scalar ratio $r$, can be expressed in terms of the $e$-fold evolution itself. These cosmological observables are defined as:
\bea
\label{eq:cosmo-observables}
& & P_s(N) = \frac{V(N)}{24\pi^2\, \epsilon(N)}, \quad n_s(N) = 1 + \frac{1}{P_s(N)} \frac{d}{dN} \,P_s(N), \\
& & \alpha_s(N) = \frac{d}{dN} n_s(N), \qquad r(N) = 16 \epsilon(N). \nonumber
\eea
All the cosmological observables are evaluated at the horizon exit $\Phi^a = \Phi^{a\ast}$ with suitable initial conditions such that one typically gets $N(\Phi^{a\ast}) \gtrsim 50$. Following the constraints from the Planck 2018 data \cite{Planck:2018jri,Planck:2018vyg}, one typically needs : $P_s \simeq 2.1 \times 10^{-9}$, $n_s=0.9651\pm 0.0044$ and $\alpha_s = - 0.0041 \pm 0.0067$. In addition, the Atacama Cosmology Telescope (ACT) data gives $n_s= 0.9666 \pm  0.0077$, Planck+ACT gives $ n_s = 0.9709 \pm  0.0038$ while Planck+ACT+DESI gives $n_s = 0.9743 \pm  0.0034$ and $\alpha_s = 0.0062 \pm 0.0052$ \cite{ACT:2025tim,ACT:2025fju,DESI:2024mwx, Frolovsky:2025iao}.

\vskip0.2cm
\noindent
{\bf Numerical Benchmark Model:} The evolutionary trajectories of various fields are determined by solving (\ref{eq:EOM2}) under the following initial conditions:
\bea
& & \Phi^a(0)=\Phi^a_0 \qquad  {\rm and} \qquad \frac{d\Phi^a}{dN}|_{N=0}=0\,.\eea
Several benchmark models compatible with observational data have been presented in \cite{Leontaris:2025hly}. For demonstration purposes, one of those models is presented as below:
\bea
\label{eq:model-M3-3-ACT}
& & \hskip-0.15cm p = 8/3, \quad \chi({\rm CY}) = -224, \quad \eta_0 = 6, \quad \sigma_0 = -4, \qquad g_s = 0.295, \\
& & \hskip-0.15cm  |W_0| = 5, \quad {\cal C}_{w_1} = 0.001, \quad  {\cal C}_{w_2} = -0.0008, \quad  {\cal C}_{w_3} = -0.0008,    \nonumber\\
& & \hskip-0.15cm {\cal C}_{w_4} = -0.1, \qquad {\cal C}_{w_5} = 0.33, \qquad  {\cal C}_{w_6} = -0.1, \qquad \lambda = - 0.00017; \nonumber\\
& & \nonumber\\
& & \hskip-0.15cm {\cal C}_{\rm up} = 5.32455, 
\quad \langle {\cal V} \rangle = 1123.23, \quad \langle t^2 \rangle = 1.14996, \quad \langle t^3 \rangle = 1.14996, \nonumber\\
& & \hskip-0.15cm \langle \phi^1 \rangle = 6.01802, \quad \langle \phi^2 \rangle = -2.41341  , \quad \langle \phi^3 \rangle = -2.41341 , \nonumber\\
& & \nonumber\\
& & \hskip-0.15cm {\cal V}^\ast = 1258.22, \quad (t^2)^\ast = 25.8, \quad  (t^3)^\ast = 25.8, \nonumber\\
& & \hskip-0.15cm \phi^{1\ast} = 6.11069, \, \, \phi^{2\ast} = 1.35 , \,\, \phi^{3\ast} = 1.35, \,\, \Delta\phi = 5.32, \,\, N = 55.5, \,\, N^\ast = 5.5,\nonumber\\
& & \hskip-0.15cm  P_s^\ast =  2.095\cdot 10^{-9}, \,\, n_s^\ast = 0.9763, \,\, \alpha_s^\ast = -5.763\cdot10^{-4}, \,\, r^\ast = 2.73\cdot10^{-3}.\nonumber
\eea

The first block of (\ref{eq:model-M3-3-ACT}) corresponds to the model dependent parameters, the second block corresponds to moduli stabiliization VEVs and the required tuning in the uplifting coefficient, while the last block presents inflationary information. The effective inflaton shift $\Delta\phi$ can be considered as the distance in the flat three-dimensional field space spanned by canonical fields ${\phi^a}$, and it simply turns out to be the Pythagorean distance between two points:
\bea
\label{eq:shiftphi-three-field}
& & \Delta \phi = \sqrt{\left(\phi^{1\ast} - \langle\phi^1\rangle\right)^2 + \left(\phi^{2\ast} - \langle\phi^2\rangle\right)^2 + \left(\phi^{3\ast} - \langle\phi^3\rangle\right)^2}~,
\eea
Here, $\phi^{a\ast}$ corresponds to the horizon exit while $\langle \phi^a\rangle$ denotes the moduli VEVs at the perturbative LVS minimum. For the benchmark model presented in (\ref{eq:model-M3-3-ACT}), we have
\begin{equation}
\Delta\phi^1 \simeq 0.0926, \qquad \Delta\phi^2 = \Delta\phi^3 \simeq 3.763 \qquad \Rightarrow \qquad \Delta\phi \simeq 5.32.
\end{equation}
Note that the first canonical field $\phi^1$ being aligned along the overall volume modulus ${\cal V}$ does not shift significantly during the inflationary process, and therefore inflation is effectively a two-field dynamics. Moreover, the key advantage of assisted inflation is demonstrated by the fact that while single-field models require $\Delta\phi \simeq 6 M_p$, the two-field approach reduces individual field excursions to be around $3.5 M_p$, mitigating concerns about trans-Planckian displacements. This also suggests that if one has sufficiently large number of `assisting' fibre moduli, one could even drive fibre inflation with sub-Planckian individual inflaton shifts.

\section{Summary and Conclusions}
\label{sec:inflaton-bound}

In this work, we have briefly reviewed (a subset of) the global model building attempts made in the context of type IIB superstring compactifications. The main focus has been to discuss models developed in the LVS framework which can be realized in two formulations; first one includes the non-perturbative effects to the superpotential while the second one uses `log-loop' effects to avoid the need of any non-perturbative contributions. 

On the broader goal of model building in string cosmology, we list the set of minimal requirements which are to be ensured to begin with. These are, for example, finding the CY threefolds with desired global properties such as Hodge numbers $(h^{1,1}, h^{2,1})$ in order to fix the number of K\"ahler moduli and complex structure moduli in the game, Euler number and second Chern numbers to quantify the $(\alpha^\prime)^3$ corrections, having the Swiss-Cheese structure through volume form etc. However, the actual demands/constraints for the global model building appear through the substructures of the CY threefolds. For example, the divisor and curve topologies have significant role on generating the effective scalar potential pieces, as various string-loop effects demand specific brane-setting following from a suitably chosen holomorphic involution. All these requirements make the model building quite a delicate process, and a systematic classification of the CY geometries with larger $h^{1,1}$ is certainly among the desirable things to do.


\section*{Acknowledgments}

We are thankful to all our collaborators for the earlier collaborations on the related subject, in particular, in the area of global model building, moduli stabilization and inflationary cosmology. PS is grateful to the organizers of the conference ``Indian String Meeting (ISM)" jointly held during Dec 09-14, 2025 at NISER+IIT-BBS, Bhubaneswar, and the organizers of the workshop ``Recent Progress in Computational String Geometry" - a BIRS-CMI event - held during January 26-31, 2026 at Chennai Mathematical Institute (CMI), Chennai, where parts of this work have been presented.  PS is thankful to the {\it Department of Science and Technology (DST), India} for the kind support.



\bibliographystyle{ws-rv-van}
\bibliography{Pramod_Shukla_CMI_Proceedings}


\end{document}